# THE HOMOGENEOUS PROPERTIES OF AUTOMATED MARKET MAKERS

Johannes Rude Jensen

University of Copenhagen, DIKU

eToroX Labs

j.jensen@di.ku.dk

Mohsen Pourpouneh

University of Copenhagen, IFRO.

mohsen@ifro.ku.dk

Kurt Nielsen

University of Copenhagen, IFRO

kun@ifro.ku.dk

Omri Ross

University of Copenhagen, DIKU

eToroX Labs

omri@di.ku.dk

**Abstract**

Automated market makers (AMM) have grown to obtain significant market share within the cryptocurrency ecosystem, resulting in a proliferation of new products pursuing exotic strategies for horizontal differentiation. Yet, their theoretical properties are curiously homogeneous when a set of basic assumptions are met. In this paper, we start by presenting a universal approach to deriving a formula for liquidity provisioning for AMMs. Next, we show that the constant function market maker and token swap market maker models are theoretically equivalent when liquidity reserves are uniform. Proceeding with an examination of AMM market microstructure, we show how non-linear price effect translates into slippage for traders and impermanent losses for liquidity providers. We proceed by showing how impermanent losses are a function of both volatility and market depth and discuss the implications of these findings within the context of the literature.

# 1    Introduction

Recent years witnessed a rapid growth in assets deposited or, 'locked', in decentralized financial (DeFi) applications deployed as smart contracts on the permissionless blockchain, Ethereum. The total value of assets under management in DeFi applications shot from a stagnant range of $400-500m to an aggregate valuation peaking at $23.3bn in late January of the following year. On multiple occasions, single DeFi applications outperformed even the most liquid centralized limit orderbook exchanges in daily transaction volumes, registering aggregate transaction volumes in excess of $60bn for the month of January 2020. At present, most assets locked in DeFi applications can be assigned within the category of smart contracts generally referred to as Automated Market Makers (AMM) further subdivided into Constant Function Market Makers (CFMM) or Token Swap Market Makers (TSMM), both a type of non-custodial smart contract implementing a deterministic pricing rule between two or more pools of tokenized assets. AMM models are composable and non-custodial, allowing external users to provide and withdraw liquidity directly to the smart contract in return for nominal trading fees and, more recently, governance token yields [1]. In contrast to the limit order book model, CFMMs process trades in constant space and time in the attempt of minimizing expensive storage operations in the distributed database as these are ultimately imposed on the end-user in the form of transaction fees paid to miners in the native unit 'gas'[2]. Though these novel applications of permissionless blockchain technologies are frequently lauded as seminal innovations within the financial industries, there is surprisingly little peer-reviewed literature on their theoretical completeness available. In this paper, we examine the common factors of AMMs and intersections between existing implementation designs by studying the commonalities and intersections between expressions of the CFMM and TSMM designs. We produce a number novel theoretical findings, proceeding into a discussion on the unique and curiously homogeneous properties of AMMs contextualizing these findings within the early results demonstrated in the literature. More specifically, we start by presenting a generalized approach by which we can derive a liquidity provision formula that satisfies a particular pricing rule. Next, we show that (ii) all AMM models deliver homogeneous results when liquidity reserves are uniform and that (iii) the price impact

of AMM trade-execution is non-linear resulting in slippage for traders and impermanent losses for liquidity providers. We proceed by showing that these losses themselves are a function of (iv) price volatility and (v) the market depth of the liquidity reserves.

## 2   Background

Given the relative conceptual novelty of AMMs, the literature is predominantly driven by practitioner and non-peer reviewed publications. AMMs present a unique set of problems in the literature, first and foremost as traders interact directly with a set of smart contracts denoting a deterministic pricing rule conditioned by available liquidity, rather than by matching counterparties through a variation of the traditional auction models designs. The original literature on 'automated' market making emerged in game theory with Hanson's seminal work on logarithmic market scoring rules [3], [4] which generated multiple novel results in the design of prediction market market makers, while never adequately solving the problem of liquidity sensitive price discovery [5].

### 2.1   Peer to Peer Exchange with Blockchain Technology

The emergence of the first generation of blockchain specific 'automated' market makers was motivated by the inefficiencies in previously dominant models for decentralized exchange, imitating the conventional central limit order book (CLOB) design. While early implementations successfully demonstrated the feasibility of executing decentralized exchange of assets on permissionless blockchain technology proved infeasible at scale. First, in the unique cost structure of the blockchain based virtual machine format [2], traders engaging with an application, pay fees corresponding to the complexity of the computation and the amount of storage required for the operation they wish to compute. Because the virtual machine is replicated on all active nodes, storing even small amounts of data is exceedingly expensive. Combined with a complex matching logic required to maintain a liquid orderbook, computing fees rapidly exceeded users' willingness to trade. Second, as 'miners' pick transactions for inclusion in the next block by the amount of gas attached to the block, it is possible to front-run state changes to the decentralized orderbook by attaching a large computational fee to a transaction including a trade, which preemptively exploits the next state change of the orderbook, thus profiting through arbitrage on a deterministic future state [6]. Subsequent iterations of decentralized exchanges addressed these issues by storing the state of the orderbook separately, using the blockchain only to compute the final settlement [7]. Nevertheless, problems with settlement frequency persisted, as these implementations introduced complex coordination problems between orderbook storage providers, presenting additional risk vectors to storage security.

## 3 Automated Market Making on the Blockchain

Responding to the shortcomings of the established methods for decentralized exchange, the latest generation of AMMs presents a new approach to blockchain enabled market design. By pooling available liquidity in trading pairs or groups in the P2C model, AMMs eliminate the need for the presence of buyers and sellers at the same time, facilitating relatively seamless trade execution without compromising the deterministic integrity of the computational environment afforded by the blockchain. While the primary context for the formal literature on blockchain based AMM has been provided by Angeris and Chitra et al. [8]–[10] the field has attracted new work on adjacent topics such as liquidity provisioning [11]–[13] and token weighted voting systems [1].

### 3.1 Stakeholder classification

AMM users can be adequately sorted in three classes by the utility they derive from the smart contracts: (i) *Traders* utilize the protocol to exchange tokenized assets seeking the best price (ii) *liquidity providers* (LPs) allocate assets to the smart contracts in return for nominal trading fees and governance token yields and, (iii) *arbitrageurs* exploit naturally occurring pricing inefficiencies between spot prices in AMMs and liquid centralized limit orderbooks (CLOB). Liquidity provisioning is computed through the issuance of 'LP shares', themselves tokenized assets representing a proportional share of the liquidity pool and the residual fees generated through trading activities[1]. While seemingly a simple solution to computing the notional value of liquidity pool fragment, LP shares have been shown to present a complex payoff function analogous to the structural properties of a perpetual option contract on the underlying segment of the liquidity pool [14]. Notably, as LP shares are themselves fungible tokens traded on adjacent AMMs they introduce a number of systemic implications to the tightly integrated single market that is the Ethereum blockchain [15].

## 3.2 Constant Function Market Makers

Constant function market makers (CFMM) denote a variation of a function in which the content of two or more liquidity pools in a trading pair or pool must approximate a constant $k$. There are three dominant implementations of constant function models, all of which can be considered expressions of the same design, holding constant the *product*, the *sum*, or the *mean* of all trading pairs or groups. In all models, assets provided by liquidity providers are pooled in open smart contracts. A trading pair consists of two or more complimentary pools of crypto assets, the total value of which is the product of the balance of the pools. Traders execute 'buy' or 'sell' orders by submitting an amount of asset $\alpha$ which returns a given amount of asset $\beta$, effectively changing the balance between the pools and, hence, the total value of the trading pair or group. The objective of the smart contract is to ensure that the total value of the liquidity pool approximates the same before and after each transaction. The constant product (1) and constant sum (2) models can be expressed as:

$$(B_\alpha - \Delta_\beta) \times (B_\beta + \gamma \Delta_\beta) = k \quad (1) \qquad (B_\alpha - \Delta_\beta) + (B_\beta + \gamma \Delta_\beta) = k \quad (2)$$

---

[1] It is shown by [11] that the return on capital accumulated by liquidity providers is proportional to the square root of the exchange rate between the two assets.

Let $α$ and $β$ be two tokenized assets, and let the pools consist of $B_α > 0$ and $B_β > 0$ token, as the balances of each pool. The value of the contract is defined as $B_α × B_β = k$. A correctly formatted transaction sends $Δ_β > 0$ of the $β$ tokens to the contract and receives $Δ_α > 0$ of $α$ tokens, effectively 'buying' $α$ tokens from the trading pair. As part of this process, an asymmetrical percentage fee $(1 − γ)$ is deducted from the incoming asset $α$ prior to the curvature, thus reducing the aggregate output in $β$ by the fee value denominated in the incoming asset [9]. Equation (1) is popularly expressed as $x × y = k$ forming a hyperbola if plotted on two axes (Figure 1.). This means that the relative asset prices approach infinity as liquidity is exhausted in one end of the pool, effectively making it impossible to empty a pool as arbitrageurs would immediately seek to profit by purchasing discounted assets in the AMM and offsetting the inventory in a CLOB market, thus returning the pool to its equilibrium state. In contrast, equation (2) denotes a straight line if plotted on two axes, expressed as $x + y = k$ which mitigates slippage but immediately exhausts the liquidity pool as any arbitrageur would be able to profit from spot price deviations by purchasing the entire pool without slippage. For this reason, implementations approximating the constant sum model use parameterized derivations of the constant product model to mitigate slippage while still making it impossible to fully empty the liquidity pool [16]. Hybrid models derived from the constant sum and constant product models are attractive choices for liquidity providers hedging volatility by holding tokenized fiat currencies, as the implied losses are reduced in non-volatile market conditions. The constant *mean* model is an expression of the constant product model in which weighted pools of multiple tokenized assets enables traders to trade in and out of up to eight assets in a single liquidity pool [17]. In a pool of three tokenized assets, this can be expressed as $(x \times y \times z)^{\frac{1}{3}} = k$, $k$ being the geometric mean of the product of the pool value.

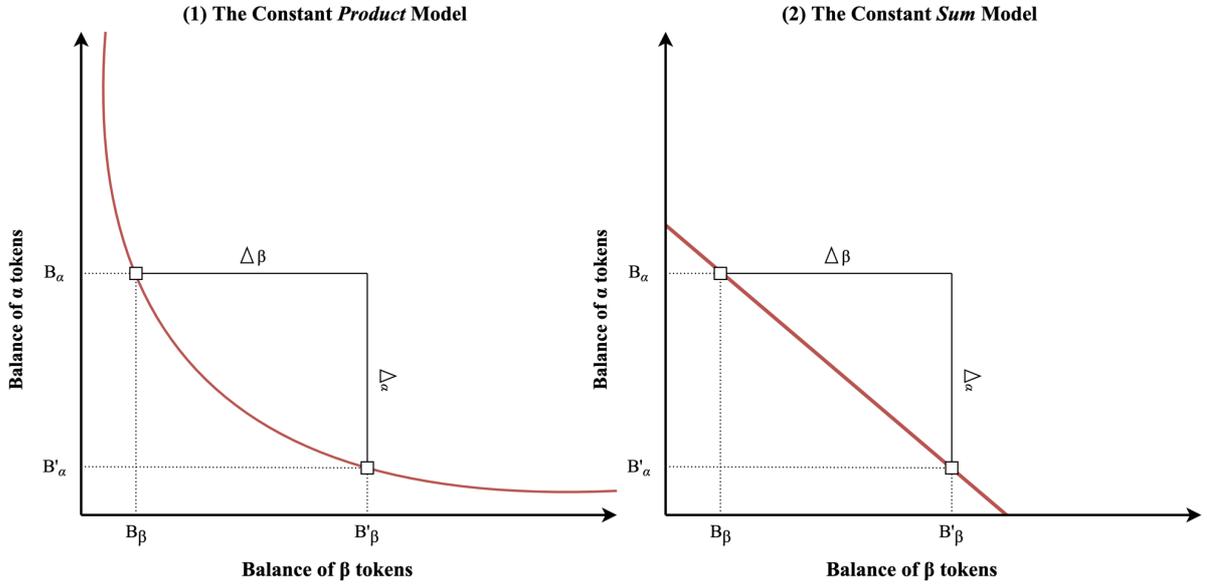

*Figure 1.*　　　　*Two expressions of the constant function automated market maker model*

### 3.3　　The Token Swap Model

The token swap (TS) model denotes a trading pair of two liquidity pools [18]. Swapping between the two tokens is facilitated through an intermediary token. Formally, let $\alpha$ and $\beta$ be two tokenized assets. A liquidity provider produces a "swap token" $\alpha\beta$, containing a reserve of $\alpha$ and $\beta$ tokens. The $\alpha\beta$ tokens can be converted to either $\alpha$ or $\beta$ tokens, based on a pricing formula derived from the balance of $\alpha$ and $\beta$ tokens in the reserve of $\alpha\beta$ tokens. Consequently, the intermediary $\alpha\beta$ token acts as a decentralized exchange trading between its two reserves. This process only depends on the balance of $\alpha$ and $\beta$ in the reserve of $\alpha\beta$ tokens, and hence the price is discovered without a need for any other parties. Let $\alpha\beta$ tokens be supported by $B_\alpha$ of $\alpha$ tokens and $B_\beta$ of $\beta$ tokens and $S_{\alpha\beta}$ be the total supply of $\alpha\beta$ tokens. Let $P_\alpha^{\alpha\beta}(P_\beta^{\alpha\beta})$ be the price of each unit of $\alpha\beta$ tokens in terms of $\alpha$ ($\beta$) tokens. At any point in time, each of the $\alpha\beta$ tokens must maintain a ratio between their total value (Supply × Price) with their reserve. This ratio is called *reserve ratio,* which is denoted by $RR_\alpha$ ($RR_\beta$) for $\alpha$ ($\beta$) tokens (such that $RR_\alpha + RR_\beta = 100\%$.), i.e., $RR_\alpha = \frac{B_\alpha}{S_{\alpha\beta} \times P_\alpha^{\alpha\beta}}$. By fixing the reserve ratio, the price of each $\alpha\beta$ token can be determined by either of its reserves. Therefore, the price of each $\alpha\beta$ token in terms of $\alpha$ tokens is:

$$P_\alpha^{\alpha\beta} = \frac{B_\alpha}{S_{\alpha\beta} \times RR_\alpha}. \tag{2}$$

From this setup, two formulas can be derived; one for calculating the amount of $\alpha\beta$ tokens that can be received by paying either $\alpha$ or $\beta$ tokens, and another one for calculating the amount of $\alpha$ or $\beta$ tokens that can be received by paying $\alpha\beta$ tokens. Formally, a transaction trading $\Delta_\beta$ of $\beta$ tokens for $\Delta_\alpha$ of $\alpha$ tokens is achieved by first converting the $\Delta_\beta$ to $\alpha\beta$ tokens and then converting the $\alpha\beta$ tokens to $\alpha$ tokens using the following formulas:

$$\# \alpha\beta \text{ tokens} = S_{\alpha\beta} \times \left(\left(1 + \frac{\Delta_\beta}{B_\beta}\right)^{RR_\beta} - 1\right) \tag{3}$$

$$\Delta_\alpha = B_\alpha \times \left(1 - \left(1 - \frac{\# \alpha\beta \text{ tokens}}{S_{\alpha\beta} + \# \alpha\beta \text{ tokens}}\right)^{\frac{1}{RR_\alpha}}\right) \tag{4}$$

The new balances after this transaction will be $B'_\alpha = B_\alpha - \Delta_\alpha$ and $B'_\beta = B_\beta + \Delta_\beta$ thus completing the trade between the two tokenized assets.

# 4     The homogeneous properties of automated market makers

Despite the hundreds of live implementations available, AMMs exhibit curiously homogeneous properties when certain basic assumptions are met. In this section, we present five propositions on the properties of AMMs.

## 4.1     A universal approach to liquidity provisioning

We can show that there exists a general approach by which we can build liquidity provision formula that satisfies a particular pricing rule.

**Proposition 1:** It is feasible to construct a liquidity provisioning formula that satisfies a particular pricing rule, using a generalizable approach.

*Proof.* In a trading pair, we denote token $x$ and $y$. As we saw above, selling a given amount of $x$ tokens to any AMM model increases the reserve by that amount, say $dx$, and decreases the reserve of $y$ by an arbitrary negative amount $dy$. Hence, the ratio $\frac{-dy}{dx}$ shows the price at which the traders has sold the $x$ tokens. This ratio, for an infinitesimally small $dx$, gives us the price offered by the AMM. Therefore, the price as a function of $x$ denominated in $y$ is given by the following simple equation:

$$p_x^y = -\frac{dy}{dx} \qquad (6)$$

Equipped with equation 6, we can define the relation between price, balances, and balance ratios for token $x, y$ and find an invariant by solving the differential equation. As an example, consider the following problem set: (i) find a curve which always offers the constant price of 1 and (ii) define a function by which the price of $x$ is based only on a ratio between $x$ and $y$ tokens in a trading pair with two liquidity pools of equal value. First, by applying equation 6 to problem (i) we get:

$$1 = -\frac{dy}{dx} \Rightarrow \int dx = -\int dy \Rightarrow x + c = -y + c'$$

$$y + x = -c + c' = k$$

Or in other words, the constant *sum* model. Second, applying equation 6 to problem (ii) implies that $p_x^y \cdot x = y$ and hence $p_x^y = \frac{y}{x}$. Here, the differential equation is:

$$\frac{y}{x} = -\frac{dy}{dx} \Rightarrow \int \frac{y}{dy} = -\int \frac{x}{dx} \Rightarrow \ln y + c = -\ln x + c' \Rightarrow \ln y + \ln x = c''$$

Rising both side to the power of $e$ and defining the $e^{c''}$ as constant $k$, we have $y \cdot x = k$ giving us the constant *product* model. A similar approach will show that the constant *mean* model follows the same structure, as it is an expression of the constant product model.

Note that the value of the constant term $k$ is determined when the pool liquidity is added to the pools. In the following proposition (2) we show why these findings hold for the TS model under the assumption of uniform liquidity reserves.

## 4.2  AMM models are homogeneous when liquidity reserves are uniform.

While market participants appear to highly persuadable by horizontal differentiation across the product range and rapidly shifting market sentiment, the properties of CPMM and TS AMMs are curiously homogeneous when reasonably common conditions are met. Extending the findings of [9] we can see that when reserves are uniform, both categories deliver equivalent results.

**Proposition 2:** The constant function model and the swap token model are equivalent when the reserve $R$ in the two juxtaposed liquidity reserves of a trading pair is uniform, formally: $RR_\alpha = RR_\beta = 50\%$.

*Proof.* To ease the notation in what follows we assume the fee to be zero i.e., $\gamma = 1$. Plugging Equation (3) into Equation (4) yields:

$$\Delta_\alpha = B_\alpha - \frac{B_\alpha}{\left(\frac{B_\beta + \Delta_\beta}{B_\beta}\right)^{\frac{RR_\beta}{RR_\alpha}}}.$$

Since $RR_\alpha = RR_\beta$, we have $\Delta_\alpha = B_\alpha - \frac{B_\alpha B_\beta}{B_\beta + \Delta_\beta}$, which implies $B_\alpha B_\beta = (B_\alpha - \Delta_\alpha)(B_\beta + \Delta_\beta)$. This is the equivalent of Equation (1), the constant product model. The marginal price of $\alpha$ tokens is defined as the price of an infinitesimally small trade. Therefore, the price can be calculated as a derivative of the formula:

$$\frac{\partial}{\partial \Delta_\alpha}\left((B_\alpha - \Delta_\alpha)(B_\beta + \gamma\Delta_\beta)\right) = 0 \implies \frac{\partial \Delta_\beta}{\partial \Delta_\alpha}\bigg|_{\Delta_\alpha=0} = \frac{1}{\gamma}\frac{B_\beta}{B_\alpha} = \frac{1}{\gamma}\frac{k}{(B_\alpha)^2} \quad (5)$$

Consequently, the price of $\alpha$ tokens (based on $\beta$ tokens) based on the available liquidity forms a hyperbola when plotted on two axis of token reserve quantities as depicted in Figure. 1. Extrapolating proposition 2, we can treat the two categories as equivalent under the given conditions, which means that the following propositions will apply equally to other models under certain conditions.

## 4.3  The price impact of AMM trade-execution is non-linear

As presented in Figure 2. the price impact of utilizing any AMM model is defined by the curvature of the function defining the rate of exchange between two or more assets. Consequently, the price impact of a given trade is equivalent to a continuous function which is tightly related to the curvature of the model. We can quantify price sensitivity for AMM trade execution simply as follows:

**Proposition 3**. Let $0 \leq f \leq 100$. Holding pool values constant for the duration of a trade, buying $fB_\alpha$ percent from a pool results in $(\frac{1}{(1-f)^2} - 1) \times 100$ percent price increase.

*Proof.* Consider selling $fB_\alpha$ of $\alpha$ tokens to the contract. By Equation (5) the price when the balance is $B_\alpha$ equals $\frac{k}{B_\alpha^2}$ and after the trade the price equals $\frac{k}{(B_\alpha - fB_\alpha)^2}$. Therefore, the percentage change in the price is:

$$\frac{P'^\alpha_\beta - P^\alpha_\beta}{P^\alpha_\beta} = \frac{\frac{k}{(B_\alpha - fB_\alpha)^2} - \frac{k}{B_\alpha^2}}{\frac{k}{B_\alpha^2}} \times 100 = \left(\frac{1}{(1-f)^2} - 1\right) \times 100.$$

Thus, the price change for selling $fB_\alpha$ percent from the pool implies a $(1 - \frac{1}{(1+f)^2}) \times 100$ percent change in the price. As we shall see below, price impact translates into *slippage* for traders and *impermanent losses*, for liquidity providers.

### 4.4 Impermanent losses are a function of price volatility

As illustrated in Figure 2. the price of an asset in both AMM categories is the product of the reserves in a trading pair. Arbitrageurs are incentivized to exploit price differences between AMM and CLOB markets by buying and selling assets from the smart contract, levelling the price, and returning the model to the equilibrium state. While this design preserves stability and retains liquidity, the cost of arbitrage equates to *impermanent* losses for liquidity providers relative to the true market price utilized by the arbitrageurs. The losses are *impermanent*, because the losses are only realized if the liquidity provider withdraws liquidity during price volatility. We can demonstrate the loss endured by the liquidity providers in both AMM models as follows:

**Proposition 4.** If $\xi > 0$ is such that $P'^\alpha_\beta = \xi \times P^\alpha_\beta$ the percentage loss resulting from providing liquidity to the pool is equivalent to $(\frac{\sqrt{\gamma\xi} + \sqrt{\frac{\xi}{\gamma}}}{1+\xi} - 1) \times 100$.

*Proof.* Let $B'_\alpha = B_\alpha - \Delta_\alpha$ and $B'_\beta = B_\beta + \Delta_\beta$, be the new balances. By equation (5), the $P'^\alpha_\beta = \frac{1}{\gamma}\frac{k}{B'^2_\alpha}$. As $P'^\alpha_\beta = \xi \times P^\alpha_\beta$ then $\frac{1}{\gamma}\frac{k}{B'^2_\alpha} = \xi \times \frac{k}{B^2_\alpha}$, since $B'_\alpha > 0$ and $B_\alpha > 0$ this implies $B'_\alpha = \frac{B_\alpha}{\sqrt{\gamma\xi}}$.

By equation (1) for the constant product model $B'_\alpha B'_\beta = B_\alpha B_\beta$ has $B'_\beta = \sqrt{\gamma\xi} B_\beta$.

The total value of the pool (in $\beta$ tokens), after arbitrageurs have made changes to the price, is $T' = B'_\alpha \times P'^\alpha_\beta + B'_\beta$. On the other hand, in the case of holding the assets, the total value of the pool would be $T = B_\alpha \times P'^\alpha_\beta + B_\beta$. This results in $\frac{T'-T}{T} \times 100 = \left(\frac{\sqrt{\gamma\xi} + \sqrt{\frac{\xi}{\gamma}}}{1+\xi} - 1\right) \times 100$.

Thus, any change in the price of either assets results in impermanent losses for the liquidity providers, which are realized if the liquidity provider withdraws their assets during a period of volatility. A closer look reveals that this is not only due to the price difference between the two markets. In fact, these results can be generalized to any transaction that is conducted with AMMs. Any transaction with the liquidity pool will affect the relative balance of tokens held in the contract, which in turn affects the price of the tokens. In markets with low volatility, impermanent losses are lower, which may result in greater profit for liquidity providers in the short term, assuming a linear accumulation of transaction fees.

## 4.5 Impermanent losses are a function of market depth

Extending proposition 4, we can show that the depth of the reserve in any token is inversely correlated to the impermanent losses suffered by liquidity providers.

*Proposition 5.* Let $0 \leq f \leq 100$. Buying $fB_\alpha$ percent from the pool results in $f \times 100$ less $\alpha$ tokens compared to an infinitely deep pool.

*Proof.* Consider buying $fB_\alpha$ of $\alpha$ tokens from the contract. This will shift the balance of $\alpha$ tokens from $B_\alpha$ to $B_\alpha - fB_\alpha$. The total amount paid by the trader is the area underneath the price curve

$$\int_{B_\alpha - fB_\alpha}^{B_\alpha} \frac{k}{x^2} dx = \frac{k}{B_\alpha - fB_\alpha} - \frac{k}{B_\alpha} = \frac{kf}{B_\alpha(1-f)}$$

Note that if the market was deep enough to mitigate slippage, buying the same amount $B_\alpha \left(\frac{f}{1-f}\right)$ of $\alpha$ tokens. Therefore, the percentage of loss from purchasing with the contract is:

$$\frac{B_\alpha \left(\frac{f}{1-f}\right) - fB_\alpha}{B_\alpha \left(\frac{f}{1-f}\right)} \times 100 = (1 - (1-f)) \times 100.$$

Note that the average price for each $\alpha$ token would be $\frac{kf}{B_\alpha^2(1-f)} = \frac{B_\beta}{1-f}$. Using a similar argument, we can show that selling $fB_\alpha$ percent to the pool results in $(1 - \frac{1}{1+f}) \times 100$ less $\beta$ tokens in comparison to an infinitely deep pool.

## 5    Discussion

AMMs are sophisticated, yet strikingly simple solutions to the problem of computing decentralized exchange subject to the computational constraints imposed by permissionless blockchain technologies. Application designers are tasked with striking the delicate balance between solving for liquidity provider incentives given by transaction fees and subsidized governance token yield, while optimizing transaction fees over price impact for the rational trader. As demonstrated above, arbitrage opportunities are a necessary but inherently loss-making product of the mechanism, resulting in impermanent losses for liquidity providers. The open-source environment in which these applications have emerged is characterized by insignificant switching costs for consumers and low entry barriers for competitors who habitually duplicate and adapt code from competing applications, a practice colloquially referred to as 'forking' [19]. These practices have produced a hostile and hypercompetitive market environment in which application designers compete to attract trading volume by offering best-price execution which, as demonstrated in proposition 3., is achieved by attracting deep liquidity reserves through ample governance token and fee distributions. In the light of the findings presented here and the unique level of transparency in the AMM markets, where traders are theoretically able to access *apriori* information on trades immediately prior to settlement [6], one may reasonably assume that these markets are characterized by symmetrical information flows and a high degree of stakeholder coordination. Yet, as suggested by [11], liquidity providers appear to exhibit behaviour tangent to strategic substitution, resulting in an adverse selection problem. The perhaps most illustrative example of this phenomenon was the so-called 'vampire attack' in which the emergent Uniswap competitor 'Sushiswap' launched an identical copy of the Uniswap AMM, subsidizing liquidity providers with a novel governance token design [1]. Consistent with the literature on information asymmetry, [20], [21] liquidity provider behaviour does not exhibit a clearly discernible pattern of coordinated behaviour but rather an increase in trading volume and liquidity around the release of news which may provide a discretionary albeit short term advantage.

Further, the results presented in this paper indicate two relevant drawbacks to the design of contemporary AMMs. First, while constant function models provide an easy optimization problem in which arbitrageurs are incentivized to level prices by maintaining equilibrium liquidity, a deterministic pricing rule necessitates a dependency to external reference prices [22]. Due to the computational constraints, such a reference price will naturally emerge adjacent high-frequency CLOB environments. While recent contributions to the practitioner literature attempts to resolve these issues by introducing *price oracles* to the token swap model [23], it is not yet clear whether the associated risks and latency issues will fully mitigate slippage and impermanent losses. Second, as demonstrated in proposition 4, slippage is a function of market depth relative to transaction volumes in the liquidity reserves of a trading pair. To mitigate excessive slippage, AMMs require reserves several orders of magnitude greater than the daily volumes traded in an arbitrary trading pair, the result of which would inevitably be low capital efficiency. While liquidity providers may, in some cases, find liquid markets for their LP shares, these markets would introduce further systemic risks to the already integrated financial network of DeFi applications.

## 6      Conclusions and Limitations

AMMs presents a fascinating innovation and an elegantly simplistic solution to the problem of decentralized exchange. While the industry is comparably noisy, we offer a demonstration of how the theoretical properties of the CFMM and TSMM models are homogeneous under certain reasonable assumptions. Like all models, there are multiple limitations to the soundness of our results. First, we assume that all agents act strategically, which may a challenging conjecture in markets characterized by excessive risk-taking and lack of regulatory oversight. Second, this study opts for modelling application transaction fees, but does not account for transaction fees for the underlying Ethereum network, which are typically paid for by AMM users. The occasionally exorbitant transaction fees suffered by network participants are likely to exercise a non-trivial impact on the strategic agents' preferences when trading across AMMs. As well-written code can drastically reduce transaction execution cost for retail volume traders, strategic agents may rightly optimize for reduced transaction costs when selecting between otherwise homogeneous products.

*Homogeneous Properties of Automated Market Makers*[14] J. Clark, "The Replicating Portfolio of a Constant Product Market," *SSRN Electron. J.*, 2020.

[15] S. Somin, Y. Altshuler, G. Gordon, and E. Shmueli, "Network Dynamics of a Financial Ecosystem," *Sci. Rep.*, vol. 10, no. 1, pp. 1–10, 2020.

[16] M. Egorov, "StableSwap - efficient mechanism for Stablecoin liquidity How it works," pp. 1–6, 2019.

[17] A. Evans, "Liquidity Provider Returns in Geometric Mean Markets," no. June, pp. 1–19, 2020.

[18] E. Hertzog, G. Benartzi, G. Benartzi, and O. Ross, "Bancor Protocol Continuous Liquidity for Cryptographic Tokens through their Smart Contracts," 2018.

[19] J. V. Andersen and C. I. Bogusz, "Self-organizing in blockchain infrastructures: Generativity through shifting objectives and forking," *J. Assoc. Inf. Syst.*, 2019.

[20] Kyle A., "Continuous Auctions and Insider Trading," *Econometrica*, vol. 53, no. 6. pp. 1315–1336, 1985.

[21] J. Chae, "Trading volume, information asymmetry, and timing information," *J. Finance*, vol. 60, no. 1, pp. 413–442, 2005.

[22] G. Angeris and T. Chitra, "Improved Price Oracles: Constant Function Market Makers," *AFT '20 Proc. 2nd ACM Conf. Adv. Financ. Technol.*, pp. 1–29, 2020.

[23] Bancor, "The V2.1 Difference: How Bancor's V2 Update Takes AMMs to the Next Level." .17